\documentclass[fleqn]{2023SCGE}
\usepackage[toc]{multitoc}

\usepackage{amsfonts}
\usepackage{amsmath}
\usepackage{amssymb}
\usepackage{array}
\usepackage{bm}
\usepackage{bbm}
\usepackage{cases}
\usepackage{color}
\usepackage{dcolumn}
\usepackage{graphicx}
\usepackage{microtype}
\usepackage{subfigure}
\usepackage{xcolor}
\usepackage{epsfig}
\usepackage{tikz}
\usepackage{verbatim}
\usepackage{cancel}

\usepackage[utf8]{inputenc} 
\usepackage{float}
\usepackage{multirow}
\usepackage{booktabs}
\definecolor{lightgray}{gray}{0.9}
\definecolor{Amber}{rgb}{1.0, 0.75, 0.0}
\definecolor{blizzardblue}{rgb}{0.67, 0.9, 0.93}

\newcommand{\PDif}[2]{\frac{\partial #1}{\partial #2}}

\begin{document}

\ensubject{subject}

\ArticleType{Article}
\SpecialTopic{SPECIAL TOPIC: }
\Year{2024}
\Month{ }
\Vol{ }
\No{ }
\DOI{ }
\ArtNo{ }

\title{Testing Cotton gravity as dark matter substitute with weak lensing}

\author[1,2]{Geyu Mo}{} 

\author[1,2]{Qingqing Wang}{} 

\author[1,2,3]{Xin Ren}{} 

\author[1,2,4]{Weitong Yan}{} 

\author[1,2]{\\Qinxun Li}{} 

\author[5,6]{Yen Chin Ong}{ycong@yzu.edu.cn} 

\author[1,2]{Wentao Luo}{wtluo@ustc.edu.cn}

\address[1]{CAS Key Laboratory for Researches in Galaxies and Cosmology, School of Astronomy and Space Science, \\ University of Science and Technology of China, Hefei 230026, China}
\address[2]{Department of Astronomy, School of Physical Sciences, University of Science and Technology of China, Hefei 230026, China}
\address[3]{Department of Physics, Tokyo Institute of Technology, Tokyo 152-8551, Japan}
\address[4]{Universitäts-Sternwarte, Fakultät für Physik, Ludwig-Maximilians Universität München, D-81679 München, Germany}
\address[5]{Center for Gravitation and Cosmology, College of Physical Science and Technology, Yangzhou University, Yangzhou 225002, China}
\address[6]{School of Aeronautics and Astronautics, Shanghai Jiao Tong University, Shanghai 200240, China}

\AuthorCitation{G Mo, Q Wang, X Ren, et al}





\abstract{Harada proposed a modified theory of gravity called Cotton gravity, and argued that it successfully explains the rotation curves of $84$ galaxies without the need for dark matter. In this work, we use the galaxy-galaxy lensing technique to test whether the modification effect of Cotton gravity can indeed be a viable substitute for dark matter. Using the spherically symmetric solution of Cotton gravity, we obtain the deflection angle via the Gauss-Bonnet theorem and the weak lensing shear. We use five galaxy catalogs divided into 5 stellar mass bins from the Sloan Digital Sky Survey Data Release 7 (SDSS DR7), each further divided into blue star-forming galaxy and red passive galaxy sub-catalogs. We find that Cotton gravity on its own has a significant deviation from the measured galaxy-galaxy lensing signals, thus it cannot replace the role of dark matter. If we consider the combination of dark matter and Cotton gravity, the modification is tightly constrained. Our analysis also applies to other modified gravity theories whose an additional linear term appears in the Schwarzschild solution.}



\maketitle
\begin{multicols}{2}

\section{Introduction}

The late time evolution of the Universe is well described by the standard $\Lambda$CDM model, which requires a mysterious dark energy to explain the accelerating expansion of the Universe \cite{SupernovaCosmologyProject:1998vns,Brout:2022vxf}, as well as a hitherto undetected dark matter, whose existence is required to properly explain the formation of galaxy and large scale structure  \cite{Blumenthal:1984bp, Davis:1985rj, Kravtsov:2012zs} and the weak lensing signals \cite{1991MNRAS.251..600B, 1994A&ARv...5..239F, Heavens:2000ad,Bradac:2006er}. 
\Authorfootnote
\noindent
It has been almost $90$ years since dark matter was proposed by Zwicky to account for why the gravitational influence of visible baryonic matter alone falls short in explaining the observed phenomena \cite{Zwicky}. Subsequent investigations into galaxy rotation curves \cite{Bosma,Sofue} provided further compelling evidence of this enigma. Nowadays, Planck Collaboration shows that dark matter constitutes a significant portion of the matter sector and contributes to the energy density of about five times more than the baryon \cite{Planck:2018vyg}, yet its nature remains unclear \cite{Bergstrom:2000pn,Bertone:2016nfn}. 

Dark matter is required if general relativity (GR) is the correct gravity theory, and most of the matter we directly observe is baryonic matter \cite{Bertone:2016nfn}. 
However, there is always the possibility that gravity has to be modified at the cosmological scale. Perhaps gravity is much stronger in weak field regime than that in Newtonian gravity \cite{Bekenstein:2004ne,Famaey:2011kh,Sanders:2002pf}. 
Ref.~\cite{1983ApJ...270..365M} first proposed Newtonian dynamics modification as an alternative to dark matter to explain the observed flat galaxy rotation curves, with the assumption that in the limit of small acceleration $a \ll a_0$, the acceleration of particle subjected to the gravitational field is stronger than that predicted in Newtonian dynamics. Here $a_0$ is a constant which has been empirically determined to be $1.2\times 10^{-10} \text{m} \cdot \text{s}^{-2}$ \cite{Milgrom:2014usa, Li:2018tdo}.  

In recent years, a new theory of gravity, called Cotton gravity, has been proposed by Harada \cite{Harada:2021bte,Harada:2021aid}, though it has so far attracted relatively few attention compared to other modified gravity theories. 
The theory led to a third-order derivative field equation, and Harada studied its spherically symmetric solution with the assumption that the metric coefficients satisfy $g_{tt}=-1/g_{rr}$. Subsequently, ref.~\cite{Mantica:2022flg} expressed the field equation of Cotton gravity with Codazzi tensor and reduced the third-order derivative field equation to the second-order derivative one, with a first-order derivative constraint equation. However, the status of the theory remains unclear as debates ensued regarding its predictability \cite{Clement:2023tyx, Sussman:2024iwk, Clement:2024pjl, Sussman:2024qsg}. Regardless, it is crucial to further test these theories of gravity in observations, beyond that of fitting galaxy rotation curves.

With the development of large sky surveys like SDSS \cite{SDSS:2000hjo}, CFHTLenS \cite{Heymans:2012gg}, HSC \cite{Aihara:2017paw} etc., weak gravitational lensing has become one of the main tools to study the cosmos. As a method to detect foreground mass overdensity, weak lensing technique is widely used in the contexts of cosmic shear \cite{KiDS:2020suj}, galaxy-galaxy lensing \cite{Mandelbaum:2005nx} and CMB lensing \cite{Lewis:2006fu}. Through analyzing image distortion of numerous background galaxies, weak lensing can directly map the gravitational field for the foreground objects, which makes it useful in constraining modified gravity, see for example, ref.~\cite{Keeton:2005jd, Luo_2021_ApJ, Ren:2021uqb, Chen:2019ftv, Wang:2023qfm, Jiang:2024otl}.

In this article, we will use the galaxy$-$galaxy lensing method to test Cotton gravity as a dark matter substitute. This work is organized as follows: In section \ref{2} we calculate the weak lensing signals of Cotton gravity. We consider two possible scenarios: without any dark matter component, and with a dark matter component (i.e., assuming that Cotton gravity effect cannot entirely replace traditional dark matter), and compare them to the $\Lambda$CDM model. In section \ref{3} we describe the data we use and show our results.
We conclude in section \ref{4}.

\section{Theory }\label{2}

\subsection{ Cotton Gravity }

Cotton gravity was first introduced in ref.~\cite{Harada:2021bte} by Harada in 2021, in which he obtained the field equation
\begin{align}\label{CottonGravity}
    C_{\rho\mu\nu}=16\pi G \nabla_{\lambda}T^{\lambda}_{~~\rho\mu\nu},
\end{align}
where Cotton tensor $C_{\rho\mu\nu}$ and the derivative of the tensor $T^{\lambda}_{~~\rho\mu\nu}$ are given respectively by:
\begin{align}
    C_{\rho\mu\nu} &\equiv \nabla_{[\mu}R_{\nu ]\rho}-\frac16 g_{\rho[\mu}\nabla_{\nu]}R, \\
    \nabla_{\lambda}T^{\lambda}_{~~\rho\mu\nu} &\equiv \frac12 \nabla_{[\mu}T_{\nu ]\rho}-\frac16 g_{\rho[\mu}\nabla_{\nu]}T,
\end{align}
in which $R_{\mu\nu},~T_{\mu\nu}$ are Ricci tensor and the usual energy momentum tensor respectively, and $R=g^{\mu\nu}R_{\mu\nu},~T=g^{\mu\nu}T_{\mu\nu}$ their respective scalar contractions. In ref.~\cite{Mantica:2022flg}, Mantica and Molinari found an equivalent expression for the field equation of Cotton gravity, re-writing eq.~\eqref{CottonGravity} into the form that makes comparisons with GR more apparent:
\begin{align}\label{CGwithCodazzi}
    R_{\mu\nu}-\frac12 R g_{\mu\nu} = 8\pi G (T_{\mu\nu}+\mathcal{C}_{\mu\nu}-\mathcal{C}g_{\mu\nu}),
\end{align}
where $\mathcal{C}_{\mu\nu}$ is an arbitrary symmetric Codazzi tensor, satisfying
\begin{align}
    \nabla_{\mu}\mathcal{C}_{\nu\rho} = \nabla_{\nu}\mathcal{C}_{\mu\rho},
\end{align}
and $\mathcal{C} = g^{\mu\nu}\mathcal{C}_{\mu\nu}$. 
As pointed out in ref.~\cite{Sussman:2024iwk}, comparing to the original formulation \eqref{CottonGravity}, this field equation \eqref{CGwithCodazzi} avoids higher order derivative problem, and more importantly avoids losing the information of the energy momentum tensor so that the vacuum condition is the usual $T_{\mu\nu} = 0$ instead of $\nabla_{\lambda} T^{\lambda}_{~~\rho\mu\nu}=0$. 
Furthermore, the field equation incorporating the Codazzi tensor can address non-vacuum conformally flat spacetime, providing a more versatile framework compared to the original formulation. For example, Friedmann equation can be obtained from the field equation \eqref{CGwithCodazzi}, with an arbitrary function of time \cite{Sussman:2023wiw}, however this cannot be done in the original form (essentially, since the Cotton tensor necessarily identically vanishes for conformally flat spacetime). 

The static spherical symmetric solution of eq.~\eqref{CottonGravity} has been made by the original paper of Cotton gravity \cite{Harada:2021aid} (see also ref.~\cite{Sussman:2023eep}), which reads
\begin{align}\label{SphereSolution}
ds^2 =& -\left( 1-\frac{2GM}{r}+2\beta r -\frac{\Lambda}{3}r^2\right)dt^2 \notag \\
    &+\frac{1}{1-\frac{2GM}{r}+2\beta r-\frac{\Lambda}{3}r^2} dr^2
    +r^2d\theta^2+r^2\sin^2\theta d\phi^2,
\end{align}
where $\beta$ is a constant (denoted by $\gamma$ in ref.~\cite{CottonGravity84Test}). In our work, we set the speed of light $c=1$ and use $\mathrm{Mpc}^{-1}$ as the unit of $\beta$. Compared to the static spherical symmetric solution of GR, we note that the solution of Cotton gravity incorporates a linear, QCD-like term $\beta r$ \cite{Harada:2021aid}. Harada claimed that the linear term makes the gravitational field stronger at large distances so that the gravitational field of baryonic matter can confine the stars and gases far out to the edge of galaxies, and help the clustering of galaxies. In this sense the linear term can be a possible substitute for dark matter \cite{Harada:2021aid}. Harada then tested this theory with the rotation velocity curves from $84$ galaxies and claimed that Cotton gravity fits the rotation velocity curves well, without requiring any dark matter component \cite{CottonGravity84Test}. Note that $\beta$ is \emph{not} a universal constant, but varies from galaxy to galaxy.

It is worth emphasizing that Cotton gravity is not the only modified gravity theory that gives rise to a linear term $\beta r$ in the static spherically symmetric solution. Many other theories have similar solution as eq.~\eqref{SphereSolution} as well, such as conformal gravity \cite{1989ApJ...342..635M}, de Rham-Gabadadze-Tolley (dRGT) massive gravity \cite{Ghosh:2015cva,Panpanich:2019mll}, and even $f(R)$ gravity \cite{Soroushfar:2015wqa}. Some specific spherically symmetric matter distribution in GR also gives rise to the metric eq.~\eqref{SphereSolution}, e.g., in the context of generalized Kiselev black hole \cite{Kiselev:2002dx}. Ref.~\cite{Gregoris:2021plc} studied the geometrical meaning of the linear term factor $\beta$ and its relation to the radius of the photon sphere so that in principle a bound of $\beta$ can be obtained from the observations of black hole shadow. Naively since our analysis only depends on the form of the metric eq.~\eqref{SphereSolution}, the conclusion also applies to all these theories, but the readers should be aware that $\beta$ can have different origins in different theories which would affect the type of conclusions one may draw. For example, in dRGT theory $\beta$ is related to the graviton mass, and it is not the only extra term that could be included in the metric; depending on the parameter values of the theory, a ``monopole'' term $c_0$ could also be added to eq.~\eqref{SphereSolution} so that the constant term in $g_{tt}$ is $-(1+c_0)$.

On the other hand, ref.~\cite{Gogberashvili:2023wed} argued that Cotton gravity admits more general solutions even assuming the static spherically symmetric condition (see eq. (19) in ref.~\cite{Gogberashvili:2023wed}; $\beta$ is denoted $\gamma$ therein). The family of solutions comes 
with two constants $C_1$ and $C_2$. However, if one requires two conditions to hold: (1) the solution is Schwarzschild-de Sitter spacetime when the effect of modified gravity vanishes; and (2) there is no cosmological horizon when $\Lambda = 0,~\beta>0$, then the only choice is $C_1 = (48/17)GM, ~C_2 = 4\Lambda/3-16\beta^2-40\beta \Lambda GM/17$, and eq.~(19) in ref.~\cite{Gogberashvili:2023wed} reduces to eq.~\eqref{SphereSolution}. Therefore, through this paper, we only consider the solution eq.~\eqref{SphereSolution}. Further research confirms eq.~\eqref{SphereSolution} as the vacuum spherical solution of Cotton gravity \cite{Sussman:2023eep}. To further simplify the analysis, we neglect the cosmological constant $\Lambda$ because of its tiny effect on the galaxy cluster scale. 

\subsection{ Shear signals }

\begin{figure*}[htp!]
\centering
\includegraphics[width=1.5\columnwidth]{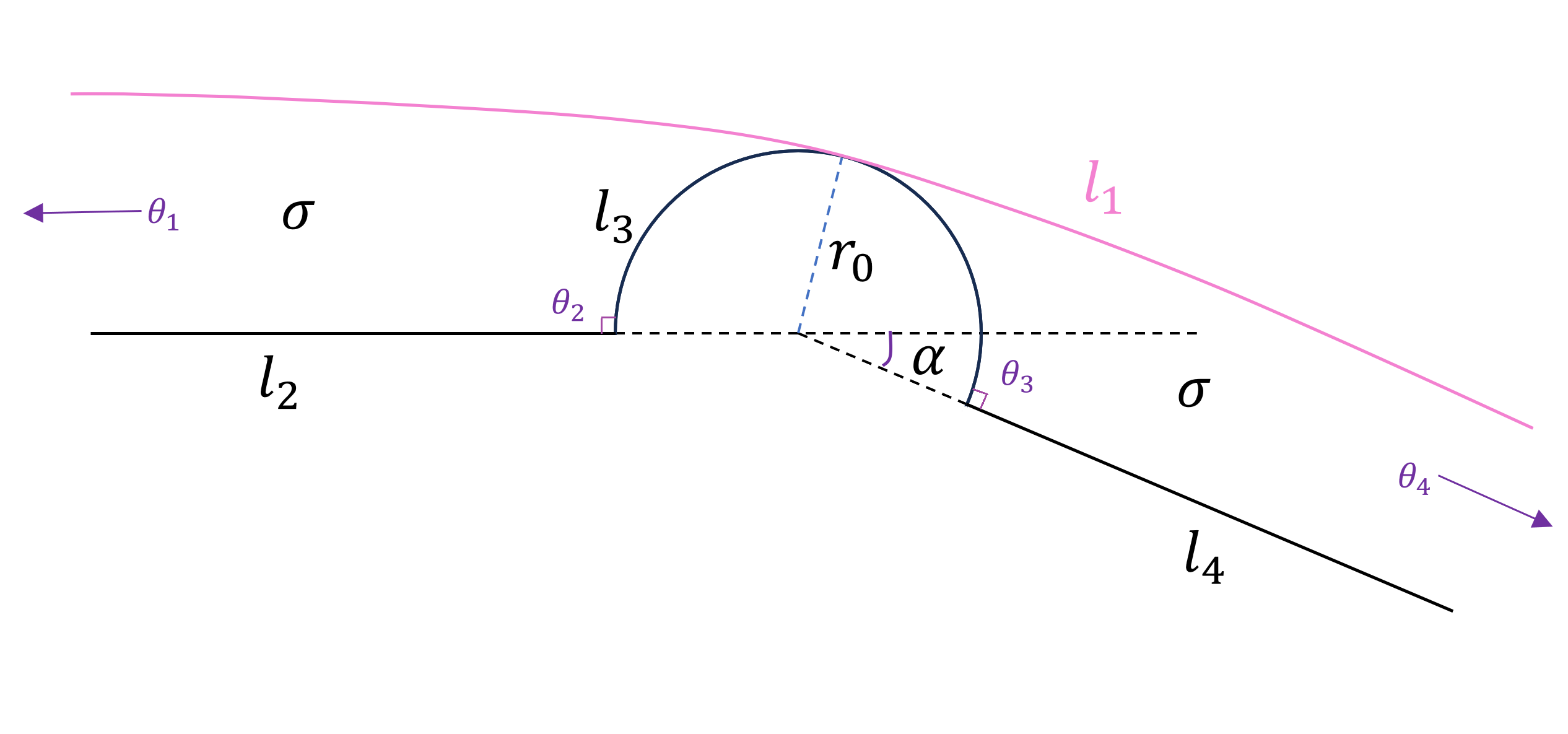}
\caption{Line segments and their bounded area to be evaluated via Gauss-Bonnet formula eq.~\eqref{AngleEq}. }
\label{Fig:Area}
\end{figure*}

The light orbit is the projection of the null geodesic on the $t=\text{const}$ hypersurfaces. Since we only consider the spherically symmetric gravitational field, we may choose the photon orbit plane as the equatorial plane $\theta=\pi/2$ without loss of generality. On the equatorial plane, we consider null geodesic, which satisfies $ds^2=0$. This condition can be used to define a Riemannian metric on the light orbit $(r,\phi)$-plane:
\begin{equation}\label{EPMetric}
    dt^2  = h_{ij}dx^i dx^j 
    = \frac{1}{\left(1-\frac{2GM}{r}+2\beta r\right)^2} dr^2+\frac{r^2}{1-\frac{2GM}{r}+2\beta r}d\phi^2.
\end{equation}
By Fermat's principle, null geodesic is also the geodesic on the light orbit plane, since 
\begin{align}
    \delta\int dt =\delta\int\sqrt{h_{ij}dx^i dx^j} = 0.
\end{align}
Following the method in ref.~\cite{Arakida:2017hrm,Takizawa:2020egm}, with eq.~\eqref{EPMetric}, the deflection angle of light rays can be obtained from the well-known Gauss-Bonnet theorem in Riemannian geometry: 
\begin{align}\label{AngleEq}
    \sum_i \theta_i+\int_\sigma K d\sigma +\int_{\partial \sigma} \kappa_g d\ell =2\pi, 
\end{align} 
where $K$ and $\kappa_g$ are the Gaussian curvature and geodesic curvature, respectively; $\theta_i$'s are the jump angles between the segments $\ell_1,~\ell_2,~\ell_3,~\ell_4$; see figure~\ref{Fig:Area}. The area $\sigma$ is also depicted in the same figure. It is straightforward to obtain the total jump angle $\sum_i\theta_i = 3\pi$. 
For the metric $h_{ij}$, the Gaussian curvature reads
\begin{align}
    K=-\frac{2GM}{r^3}+\frac{3G^2M^2}{r^4}-\frac{2GM\beta}{r^2}-\beta^2,
\end{align}
The only non-geodesic is the circular segment $\ell_3$, whose geodesic curvature is given by the integral
\begin{align}
    \int_{\partial \sigma}\kappa_g d\ell&= \int^{- \alpha}_{\pi} \sqrt{\mathrm{det}|h_{ij}|}\epsilon_{ij}v^i\frac{d v^j}{dt}dt\notag \\
    &= \int^{- \alpha}_{\pi}\frac{\left( 1-\frac{GM}{r_0}+3\beta r_0 \right)}{\sqrt{1-\frac{2GM}{r_0}+2\beta r_0}}d \phi. 
\end{align}
Neglecting the higher order terms $O(G^3M^3)$ and $O(\beta^2)$, the deflection angle of light, $\alpha$, can be deduced:
\begin{align}
    \alpha \approx& \frac{4GM}{r_0}+\left( \frac{15\pi}{4}-4 \right)\frac{G^2M^2}{r_0^2}-2(\pi-2)GM\beta\notag \\
    &+2\left( \frac{15\pi}{4}-20 \right)\frac{G^2M^2\beta}{r_0}\notag \\
    \approx& \frac{4GM}{R}+\frac{15\pi}{4}\frac{G^2M^2}{R^2}-2\pi GM\beta-2 \left(\frac{22G^2M^2\beta}{R}\right),
\end{align}
where $R$ is the radius from the gravitational center to the light ray on the lensing plane. Note that the deflection angle is still the spatial gradient of the lensing potential, yet the gravitational potential is not the Newtonian potential but the corresponding gravitational potential under Cotton gravity. Due to the linear term $\beta r$, it is difficult to derive an explicit gravitational potential, therefore we derive the deflection angle by Gauss-Bonnet theorem. We can now see that at the linear order of $\beta$, the effect of Cotton gravity reduces the deflection angle from GR's results, which implies that more mass is required if the deflection angle is to remain unchanged. At the scale of our concern, $R \sim 1 ~\mathrm{Mpc}/h$, the contribution of Cotton gravity is about $10^{-4}$ less than the Newtonian term, but about $10^{2}$ higher than the GR correction term.  

Recall that the claim of ref.~\cite{Harada:2021bte} is that the effect from the linear term $\beta r$ can replace the role of dark matter. If so, the lensing galaxies can be regarded as a point mass instead of a broad density profile in the scenarios with dark matter. We recall that the lensing shear for a spherically symmetric gravitational field can be obtained from the deflection angle \cite{Narayan,Bartelmann:2016dvf}. In the standard notations where $D_s$, $D_d$ and $D_{dS}$ denoting the distances between the observer and the source, the observer and the lens, and the lens and the source, respectively, the lensing shear is given by
\begin{align}
    \gamma &= D_d^2\sqrt{\frac{1}{4}\left( \frac{\partial^2 \psi}{\partial x^2} - \frac{\partial^2 \psi}{\partial y^2} \right)^2+\left( \frac{\partial^2 \psi}{\partial x \partial y} \right)^2} \notag \\
    &=\frac{D_d D_{ds}}{2D_s}\left( \frac{\alpha(R)}{R}-\PDif{\alpha(R)}{R} \right),
\end{align}
where $\psi$ is the lensing potential (see \cite{Bartelmann:2016dvf} for the precise definition), with $(x,y)$ denoting the standard coordinates on the two-dimensional lens plane. Finally the lensing shear is obtained to be
\begin{align}
    \gamma = \frac{D_d D_{ds}}{D_s}\left(\frac{4GM}{R^2}-\frac{\pi \beta GM}{R}-\frac{44\beta G^2M^2}{R^2}+\frac{45\pi G^2M^2}{8R^3}\right).
\end{align}
The excess surface density is given by 
\begin{align}
    \Delta \Sigma (R) = \gamma \Sigma_c = \frac{M}{\pi R^2}-\frac{\beta M}{4R}-\frac{11\beta GM^2}{\pi R^2}+\frac{45 G M^2}{32R^3},
\end{align}
where $\Sigma_c = {D_s}/{(4\pi G D_d D_{ds})}$ is the critical surface density.

On the other hand, in $\Lambda$CDM model, we describe the galaxy host dark matter halos by the Navarro$-$Frenk$-$White (NFW) profile \cite{1996A&A...313..697B, Navarro:1996gj}, which is introduced by $N-$body simulation to fit equilibrium density profile of cold dark matter halos, and independent of the halo mass and cosmological parameters \cite{Golse:2001ar}. In particular the cosmological constant $\Lambda$ does not play a role here (hence we also neglect $\Lambda$ in the Cotton gravity case). The NFW profile reads:
\begin{align}\label{NFWProfile}
    \rho (r) = \frac{\rho_c}{(r/r_s)(1+r/r_s)^2}, 
\end{align}
where $\rho_c$ is a characteristic density and $r_s$ a scale radius. For convenience, we express $\rho_c$ and $r_s$ by the halo virial mass $M_h$ and the concentration parameter $c$:
\begin{align}
    M_h=\frac{4\pi v}{3}c^3 r^3_s \rho_{\mathrm{crit}},\\
    \rho_c = \frac{v}{3}\frac{c^3}{\ln(1+c)-\frac{c}{1+c}} \rho_{\mathrm{crit}},
\end{align}
where $\rho_{\mathrm{crit}}$ is the critical density. Here we set $v=200$. From eq.~\eqref{NFWProfile} one can derive the surface density straightforwardly \cite{1996A&A...313..697B, 2000ApJ...534...34W}:
\begin{align}
    \Sigma_h (R) = \int^{+\infty}_{-\infty} \rho\left(\sqrt{R^2+z^2}\right) dz = 2\rho_cr_sF(x),
\end{align}
where $x=R/r_s$ and
\begin{align*}
    F(x) = 
    \begin{cases}
        \frac{1}{x^2-1}\left[1-\frac{1}{\sqrt{1-x^2}}\ln\left( \frac{1+\sqrt{1-x^2}}{x} \right)\right], & x<1, \\
        \frac{1}{3}, & x=1, \\
        \frac{1}{x^2-1}\left[1-\frac{1}{\sqrt{x^2-1}}\arccos\left(\frac{1}{x}\right)\right], & x>1.
    \end{cases}
\end{align*}
The average of surface density is then
\begin{align}
    \bar{\Sigma}_h(R) = \frac{1}{\pi x^2}\int^{x}_0 2\pi \Sigma(x) x ~dx = 4\rho_c r_s g(x),
\end{align}
with
\begin{align*}
    g(x) = 
    \begin{cases}
        \frac{1}{x^2}\left[\ln\frac{x}{2}+\frac{1}{\sqrt{1-x^2}}\ln\left( \frac{1+\sqrt{1-x^2}}{x} \right)\right], & x<1, \\
        1-\ln 2, & x=1,\\
        \frac{1}{x^2}\left[\ln\frac{x}{2}+\frac{1}{\sqrt{x^2-1}}\arccos\left(\frac{1}{x}\right)\right], & x>1.
    \end{cases}
\end{align*}
The excess surface density is then simply 
\begin{align}
    \Delta\Sigma_h(R)=\bar{\Sigma}_h(R)-\Sigma_h(R).
\end{align}
Including the baryonic component of the galaxy, the total excess surface density is
\begin{align}
    \Delta\Sigma_{NFW}(R) = \frac{M_b}{\pi R^2}+\Delta\Sigma_h(R), 
\end{align}
where $M_b$ is the baryonic mass.

In ref.~\cite{Harada:2021bte} the author did not assume any dark matter halo profile and explained galaxy rotation curve with the effect of modified gravity alone, but in principle 
there can still be dark matter even in this theory. 
Therefore, we also consider the NFW profile as the dark matter profile in this modified gravity framework. In this scenario, the excess surface density becomes
\begin{align}
    \Delta \Sigma (R) =& \Delta\Sigma_{NFW}(R)-\frac{\beta M_{_{NFW}}(R)}{4R}-\frac{11\beta GM_{_{NFW}}^2(R)}{\pi R^2} \notag\\
    &+\frac{45 G M_{_{NFW}}^2(R)}{32R^3},
\end{align}
where $M_{_{NFW}}$ is the mass of NFW profile in the radius $R$.

\section{Data and Result}\label{3}
We use galaxy$-$galaxy lensing signals measurement from ref.~\cite{Luo_2021_ApJ} to test the extra contribution from cotton gravity at the cluster scale. There are $400,608$ galaxies in the lensing catalog, and the effect from nearby structures has been minimized when selecting the lensing galaxies \cite{Yang:2006zf}. The galaxies have been extinction corrected, with redshift within $0.01\leq z \leq 0.2$ and the $r-$band magnitude brighter than $17.72$. The lensing galaxies are selected so that there is only one central galaxy, without any other galaxy brighter than $17.77$ in its projected virial radius and within the redshift dispersion less than the virial velocity of the dark halo. The lensing catalog is divided into two catalogs containing blue star-forming galaxies and red passive galaxies respectively. Each of these two catalogs is subdivided into five catalogs according to the stellar mass of the galaxies, following ref.~\cite{Luo_2021_ApJ}. The mean stellar mass of these ten catalogs is shown in table \ref{Result:details}. The gas is also taken into account when considering the total baryonic matter mass, which can be written as \cite{2010A&A...518L..61B,2014A&A...564A..67B}  
\begin{align}
    M_g = (1+f)M_{\star},
\end{align}
where $f$ is given by \cite{2014A&A...564A..67B}
\begin{align}
    \log (f) = -0.69\log \left(\frac{M_{\star}}{h^{-2}M_{\odot}}\right)+6.63,
\end{align}
and $M_{\star}$ denotes the stellar mass of the galaxy. On the other hand, in our lensing catalog, the total possible systematic error in $2\sigma$ of the shear is about $-0.091<\delta\gamma/\gamma<0.208$, including selection bias, point spread function reconstruction bias, point spread function dilution bias, shear responsivity error, and noise rectification bias. 

\begin{table*}[htp!]
\centering
\caption{Best-fit parameters of NFW and Cotton Gravity. Here $M_{\star}$ is the stellar mass obtained from the galaxy luminosity, $M_h$ is the mass of the dark matter halo, $c$ is the concentration parameter, and $M_b$ is the total baryonic mass \cite{Luo:2016ibp}. }
\begin{tabular}{ccccccccc}
\hline
\hline
\multirow{3}*{Catalog} & \multirow{3}*{ $\log\left(\frac{M_{\star}}{h^{-2}M_{\odot}}\right)$ }& \multicolumn{2}{c}{NFW} & \multicolumn{2}{c}{Cotton (no DM)}& \multicolumn{3}{c}{Cotton (DM)} \\ 
\cmidrule(lr){3-4}\cmidrule(lr){5-6}\cmidrule(lr){7-9} & & \multirow{2}*{$\log(\frac{M_h}{h^{-1} M_{\odot}})$ }& \multirow{2}*{$c$} & \multirow{2}*{$\log(\frac{M_b}{M_{\odot}})$} & \multirow{2}*{$\frac{10^{4}\beta}{\mathrm{Mpc}^{-1}}$} & \multirow{2}*{$\log(\frac{M_h}{h^{-1} M_{\odot}})$} & \multirow{2}*{$c$} & \multirow{2}*{$\frac{10^{4}\beta}{\mathrm{Mpc}^{-1}}$} \\
\\
\hline
Red 1 &9.918 & 11.003 & 6.252 & 12.265 &0.003 &11.003 &6.252 &13.579\\
Blue 1 &9.916 & 10.385 & 6.386 & 11.665 & 0.005 &10.385 &6.386 &1.062 \\
Red 2 &10.479 & 11.235 & 4.978 & 12.479 & 0.186 &11.235 &4.978 &0.121 \\
Blue 2 &10.479 & 11.238 & 6.180 & 12.357 & 38.932 &11.238 &6.180 &30.354 \\
Red 3 &10.679 & 11.566 & 7.411 & 12.612 & 3.271 &11.566 &7.411 &0.004 \\
Blue 3 &10.679 & 11.473 & 8.701 & 12.470 & 0.002 &11.473 &8.701 &0.286 \\
Red 4 &10.776 & 11.640 & 5.768 & 12.649 & 31.366 &11.640 &5.768 &0.029 \\
Blue 4 &10.776 & 11.815 & 5.405 & 12.734 & 0.001 &11.536 &3.949 &0.003 \\
Red 5 &10.936 & 11.768 & 7.140 & 12.843 & 0.109 &11.768 &7.140 &1.173  \\
Blue 5 &10.936 & 11.815 & 5.624 & 12.745 & 0.699 &11.815 &5.405 &0.002 \\
\hline
\end{tabular}
\label{Result:details}
\end{table*}

The source catalog is constructed by ref.~\cite{Luo:2016ibp}, containing about $40$ million galaxies from SDSS DR$7$ data, within the redshift range of $0.5\leq z \leq 1$. The shear signals $\Delta \Sigma$ are obtained from the mean of source galaxies shear \cite{Miyatake:2018lpb},
\begin{align}
    \Delta\Sigma = \frac{1}{2\bar{R}}\frac{\sum_l w_l \sum_s w_{ls} e_{T,ls}[\langle \Sigma_c^{-1} \rangle_{ls}]^{-1}}{\sum_l w_l \sum_s w_{ls}},
\end{align}
where $\bar{R}$ is the responsivity of the shear estimator, and for SDSS DR$7$ data $\bar{R}\approx 0.84$; $w_{ls}$ is the weight of every source galaxy $w_{ls}= (\langle \Sigma_c^{-1} \rangle_{ls})^2/(\sigma^2+e_{rms}^2)$, which is the function of the shape noise $\sigma$ and the shape measurement error $e_{rms}$; $w_l$ is the weight for the lensing samples. To measure the shear signals, we divide the projected radius into $6$ equal logarithmic bins in the range $0.01\mathrm{Mpc}~ h^{-1}\leq R \leq 1 \mathrm{Mpc}~ h^{-1}$.

We use the minimized $\chi^2$ optimization to fit the observational data for each model. In the NFW model, we leave the dark halo mass and the concentration parameter as free parameters, while fixing the baryonic matter mass by observational data. In the scenario of Cotton gravity without dark matter, we set the baryonic matter mass and the parameter of Cotton gravity $\beta$ as free parameters. In Cotton gravity with dark matter scenario, we let the dark halo mass, the concentration parameter and the parameter of Cotton gravity be free, while fixing the baryonic matter mass by observational data.

\begin{figure*}[htp]
\centering
\includegraphics[width=1.6\columnwidth]{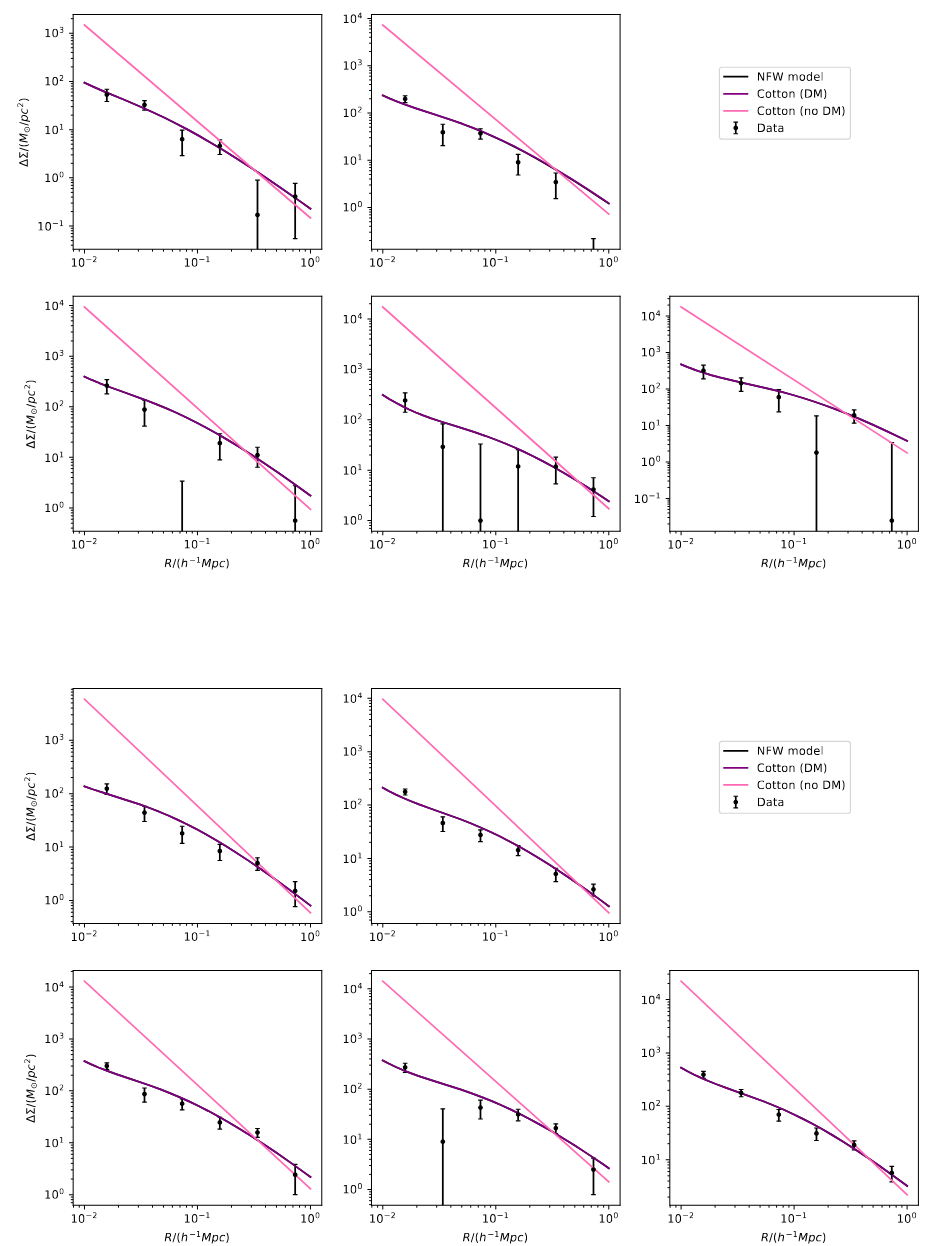}
\caption{ 
The best fit for the NFW model, Cotton gravity without dark matter scenario (Cotton no DM) and with dark matter scenario (Cotton DM) with data from the blue star forming galaxies catalog (Upper) and from the red passive galaxies catalog (Lower), respectively. }
\label{Fig:Res}
\end{figure*}

\begin{figure*}[htp]
\centering
\includegraphics[width=1.7\columnwidth]{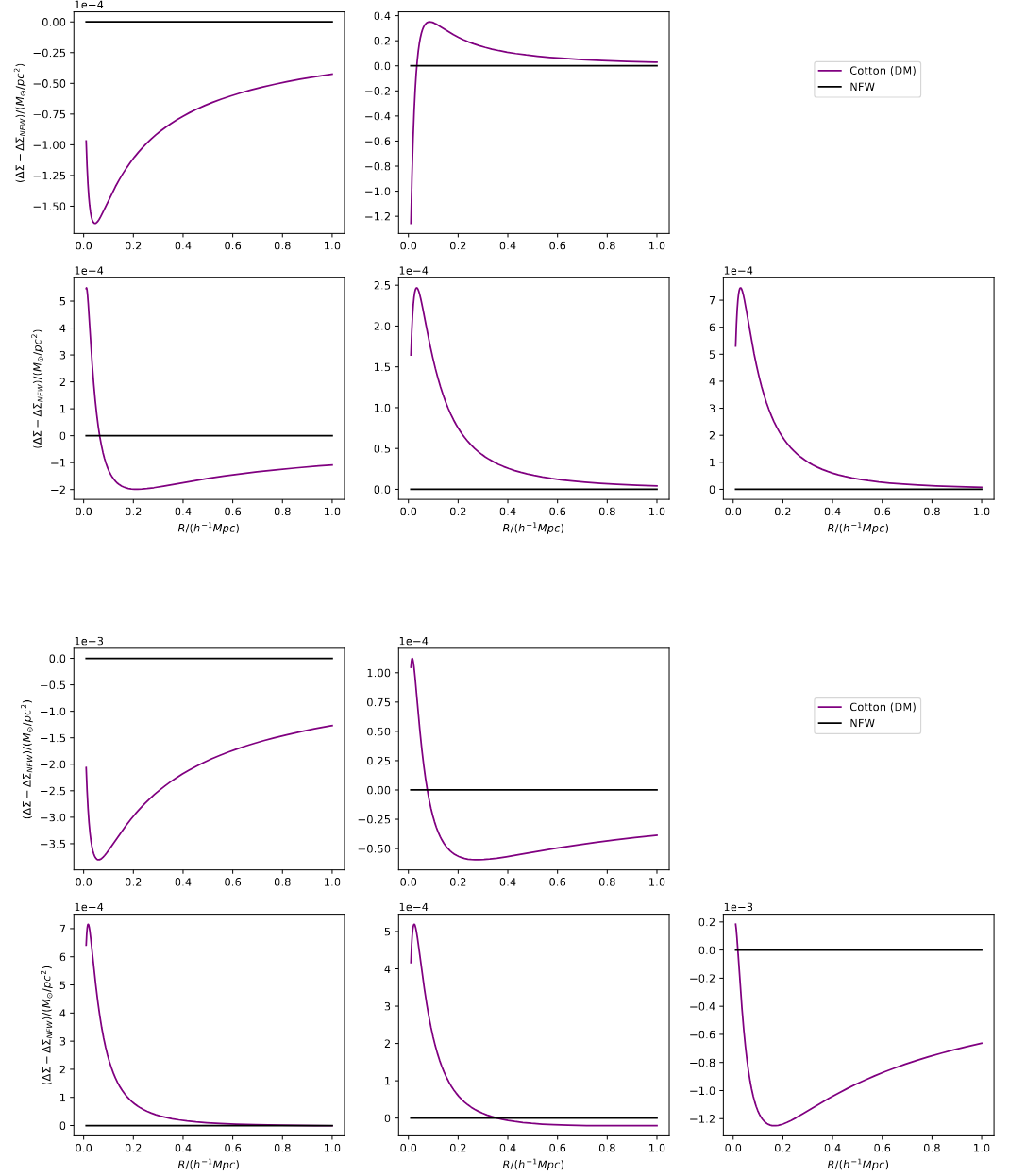}
\caption{Difference between the best fit for the NFW model and Cotton gravity with dark matter scenario of star forming galaxies catalog (Upper) and the red passive galaxies catalog (Lower), respectively. }
\label{Fig:Dif}
\end{figure*}

Best fit results of the lensing signals are shown in figure~\ref{Fig:Res} and figure~\ref{Fig:Dif} by comparing Cotton gravity and the NFW profile in $\Lambda$CDM model. The best fit of the free parameters of each model shown in table~\ref{Result:details} shows that the order of magnitude for the values of $\beta$ is consistent with the fitting results of galaxy rotation curves \cite{CottonGravity84Test}. However, it is clear from figure~\ref{Fig:Res} that the curve of Cotton gravity without dark matter scenario does \emph{not} fit the observational lensing data and the best fit baryonic matter mass is overestimated compared to the stellar mass data, indicating the requirement for dark matter even in Cotton gravity theory. Note the curves of the NFW model overlap completely with the scenario of Cotton gravity with dark matter, since the contribution of the extra term from Cotton gravity is comparatively tiny compared with the added dark matter, as mentioned in section 2 and shown in figure~\ref{Fig:Dif}. We also use MCMC method to estimate the effect of systematic error. The estimation result shows the systematic error only engenders about 0.36 dex to the estimation of the parameters, which has no significant impact on our conclusion. 

We now calculate the Bayesian information criterion (BIC) for each model with dark matter \cite{Liddle:2007fy} to compare their goodness-of-fit. The BIC is 
\begin{align}
    \text{BIC} = -2\ln(\mathcal{L})+k\ln(N),
\end{align}
where $\mathcal{L}$ denotes the likelihood, and $k,~N$ denote the number of free parameters and the sample size, respectively. The values of BIC of red and blue galaxy catalogs for each model, and the difference between NFW and Cotton Gravity are shown in table \ref{Result:BIC}. The results show that the NFW model has lower BIC values, and thus is still favored by weak lensing signals. Note that $\Delta \text{BIC} \gtrsim 10$, thus we obtained a significant disfavor for Cotton gravity model either with or without dark matter since the theory has more parameters. 

\begin{table}[H]
\centering
\caption{BIC comparison between NFW and Cotton Gravity. }
\begin{tabular}{ccccc}
\hline
\hline
\multirow{2}*{Catalog} & \multirow{2}*{Criterion} & \multirow{2}*{NFW} & \multicolumn{2}{c}{Cotton Gravity} \\ 
\cmidrule(lr){4-5}& & & no DM & with DM \\
\hline
\multirow{2}*{Red} & $\text{BIC}$ & 36.990 & 53.997 & 48.960 \\
&$\Delta \text{BIC}$ & 0.0 & 17.007 & 11.970 \\
\hline
\multirow{2}*{Blue}& $\text{BIC}$ & 44.190 & 61.198 & 54.118 \\
&$\Delta \text{BIC}$ & 0.0 & 17.008 & 9.928 \\
\hline
\end{tabular}
\label{Result:BIC}
\end{table}

\section{Conclusion: Weak Lensing Disfavors Cotton Gravity}\label{4}
In this work,
we have derived the deflection angle and the lensing shear for Cotton gravity using its spherically symmetric solution, which has an extra linear term in the metric function compared to GR. We then used galaxy$-$galaxy lensing signal measurement data from ref.~\cite{Luo_2021_ApJ,Luo:2016ibp} to test Cotton gravity. We found that compared to GR there is a small modification for the deflection angle and the lensing shear in Cotton gravity. In the scenario without dark matter, the modification term from Cotton gravity is too small to replace dark matter halo at the galaxy cluster scale, despite it works relatively well to explain the flat galaxy rotation curve of galaxies in \cite{CottonGravity84Test}.  If we add dark matter into the Cotton gravity framework,
the Bayesian information criterion suggests that at the galaxy cluster scale, Cotton gravity does not perform as well as general relativity.

There have been some important discussions about the theoretical issues of Cotton gravity in the literature recently. Ref.~\cite{Clement:2024pjl} pointed out that Cotton gravity is under-determined, with some degrees of arbitrariness hidden in the original formulation, which still persisted in the Codazzi tensor formulation, thus concluding that Cotton gravity is not a predictive theory \cite{Clement:2023tyx}. On the other hand, ref.~\cite{Sussman:2024qsg} argues that with the Codazzi formulation, the under-determination problem of Cotton gravity mentioned in ref.~\cite{Clement:2024pjl} no longer exists, and they argued that the spherical symmetry solution of Cotton gravity is still given by eq.~\eqref{SphereSolution}. Considering the ambiguous theoretical status, it is all the more important to test Cotton gravity empirically beyond the galaxy rotation curve fitting. In this work we have done precisely this, and our results do not favor Cotton gravity, at least with the simplest solution employing the linear term $\beta r$. 

Finally, we note that the spherically symmetric solution means that the center of the gravitational field is still located at the center of the mass distribution. That implies the scenario for Cotton gravity without dark matter would have difficulties explaining bullet-like clusters \cite{Bradac:2006er,Brownstein:2007sr}. With non-spherical geometry some modified Newtonian dynamics models can generate a multi-centered system while recovering the weak lensing signals of the bullet cluster $1E~ 0657-56$, even without dark matter \cite{Angus:2006qy}. Whether more complicated solutions in Cotton gravity can be constructed to better fit observational data requires more studies in the future.

\section*{Acknowledgements}
\Acknowledgements{
We are grateful to Yi-Fu Cai, Bo Wang, Chengze Dong, Dongdong Zhang, Pengbo Xia for valuable discussions.
This work is supported in part by the National Key R\&D Program of China (Grant No. 2024YFC2207500, 2021YFC2203100), CAS Young Interdisciplinary Innovation Team (JCTD-2022-20), 
Natural Science Foundation of China (Grant No. 92476203, 12433002, 12261131497, 12003029), 111 Project for ``Observational and Theoretical Research on Dark Matter and Dark Energy" (Grant No. B23042), by Fundamental Research Funds for Central Universities, by CSC Innovation Talent Funds, by USTC Fellowships for International Cooperation, and by USTC Research Funds of the Double First-Class Initiative. }

\InterestConflict{The authors declare that they have no conflict of interest.}



\end{multicols}
\end{document}